\documentclass[]{aa}

\usepackage[varg]{txfonts}
\usepackage{natbib}
\usepackage[T1]{fontenc}
\usepackage[utf8]{inputenc}
\usepackage{graphicx}

\begin{document}

\title{Advanced aspects of Galactic habitability}

\titlerunning{Advanced aspects of Galactic habitability}
\authorrunning{Đošović et al.}

\author{Vladimir Đošović\inst{1}
  \and Branislav Vukotić\inst{2}\thanks{Corresponding author} 
  \and Milan M. Ćirković\inst{2}} 


\institute{Department of Astronomy, Faculty of Mathematics, University of Belgrade, Studentski Trg 16, 11000 Belgrade, Serbia, \email{vladimir{\_}djosovic@matf.bg.ac.rs}
  \and Astronomical Observatory, Volgina 7, P.O.Box 74 11060 Belgrade, Serbia, \email{bvukotic@aob.rs}, \email{mcirkovic@aob.rs}} 

\date{Received \today / Accepted }

\abstract {Astrobiological evolution of the Milky Way (or the shape of its `astrobiological landscape') has emerged as a key research topic in recent years. In order to build precise, quantitative models of the Galactic habitability, we need to account for  two opposing tendencies of life and intelligence in the most general context: the tendency to spread to all available ecological niches (conventionally dubbed `colonization') and the tendency to succumb to various types of existential catastrophes (`catastrophism'). These evolutionary tendencies have become objects of study in fields such as ecology, macroevolution, risk analysis, and futures studies, though a serious astrobiological treatment has so far been lacking.} 
{Our aim is to numerically investigate the dynamics of opposed processes of expansion (panspermia, colonization) and extinction (catastrophic mechanisms) of life in the Galaxy. } 
{We employed a new type of numerical simulation based on 1D probabilistic cellular automaton with very high temporal resolution, in order to study astrobiological  dynamics.}   
{While the largest part of the examined parameter space shows very low habitability values, as expected, the remaining part has some observationally appealing  features that imply, among other things, a reduction in the amount of fine-tuning necessary for resolving the Fermi paradox.}    
{Advanced aspects of Galactic habitability are amenable to precision studies using massive parallel computer simulations. There are regions of the parameter space that correspond to a quasi-stationary state satisfying observable constraints and possessing viable SETI targets.} 

\keywords{methods: numerical - Galaxy: evolution - extraterrestrial intelligence - astrobiology}
\maketitle

\section{Introduction: Habitability and colonization}

Studies of habitability -- circumstellar, galactic, and cosmological --  have rapidly become the main topic of astrobiological research in recent years \citep[e.g.][]{2005ARA&A..43...31C,2010A&ARv..18..383J,2010IJAsB...9..273H,2012AREPS..40..597L,2015ApJ...810L...2D}. Both important theoretical insights and novel numerical models have been developed and deployed in the course of the last decade in order to obtain a better quantitative handle on the hitherto mostly intuitive concept of habitability. In particular, cosmological structure formation simulations \citep{2016MNRAS.459.3512V,2017IJAsB..16...60F} and semi-analytic merger trees \citep{2018MNRAS.475.1829S} have been used successfully to that effect. Clearly, there is much work to be done in this area, with refinement of both computing resolution and empirical constraints on habitability. 

There is, however, another aspect of habitability that has not yet been quantitatively studied. This is the interplay between (often intuitively assumed, but rarely explored in detail) two generalized ecological tendencies that characterize life irrespective of its exact definition: the tendency to spread to all available niches and the tendency to become extinct as a consequence of catastrophic changes in its physical environment. While intuitively generalized from our observations of terrestrial life, there is no reason whatsoever to assume that extraterrestrial life, if it exists, will be exempt from either of these two grand trends. The spread of life over various cosmic distances has been usually operationalized as either panspermia or intentional colonization by intelligent life.  Since the classical study of \citet{1981Icar...46..293N}, there have been several specific studies of the modes of spreading of intelligent life through the Galaxy \citep[e.g.][]{bjoerk2007exploring,cotta2009computational,starling2014virulence}. On the other hand, most of the contemporary studies of habitability take into account at least some of the astrophysical catastrophic risks, such as those from supernovae and gamma-ray bursts \citep[e.g.][]{1995PNAS...92..235E,2002ApJ...566..723S,2007IJAsB...6...19G,2008OLEB...38..535C,2016ApJ...832...38G,2018MNRAS.475.1829S}. In fact, one could even state that the `catastrophic' aspects of the overall astrobiological evolution have been overemphasized in recent years. The very concept of habitable zones, bounded in space and time, which has emerged as the `bread and butter' of astrobiology \citep[e.g.][]{2005OLEB...35..555G}, testifies to the resurgence of catastrophism as a key ingredient in any viable astrobiological theory. 

In contrast, the aspect of astrobiological evolution that deals with spreading of life has thus far been largely neglected or downplayed in quantitative models \citep[even though the philosophical justification of life's tendency to spread and consequent implications for astrobiology is well-studied under the title of the  continuity thesis; cf.][]{fry2000emergence,cirkovic2012astrobiological}. Often a lip service is paid to the tendency of life to spread and fill all available ecological niches \citep[e.g.][]{hanson1998burning,sagan2000carl}, but it is seldom analysed and modelled in detail. There are essentially two modes of spreading of life which might be co-present in the overall astrobiological landscape: panspermia (simple life) and colonization (complex life, specifically technological civilizations). The two are not entirely disjunctive, since the emergence of spacefaring civilization is accompanied by possibility of {\bf directed panspermia} \citep{crick1973directed,sleator2017directed}. Actual concerns about planetary contamination and planetary protection \citep[e.g.][]{rummel2001planetary} testify how even a primitive technological civilization, like the present-day human one, faces this problem quite early in its spacefaring attempts. Therefore, it makes sense to discuss both these aspects of the same `spreading of life' tendency in a unified way, as we are doing in the present model; we refer to this dual tendency as `colonization', while keeping in mind that the model parameters are actually adaptable enough to include all kinds of panspermia.  (A similar dual construal is applicable on the catastrophism side of the story, where we can subsume both naturally occurring and technogenic events, such as the nuclear winter, which is capable of causing widespread ecological devastation and extinction, under the same controlling parameter.)

Therefore, trying to achieve a more complete picture is highly justified.  Cellular automata approach, a grid of cells (each described with some variable called state) that evolve in discrete time steps \citep[see][]{Ilachinski2001} has been shown to be a promising modelling platform for such phenomena \citep[][]{2012OLEB...42..347V,2012POBeo..91..221V}. However, since life-friendly matter conditions  are usually found on planetary bodies, and in particular the habitability of systems of such bodies is greatly influenced with galactic parameters, the required spatio-temporal dynamical range for such models is huge. Consequently, vast computational resources are required. Clearly, a practical difficulty in incorporating these two opposing tendencies in a single synthetic model  is the discrepancy between relevant timescales in the general case. The present study seeks to address at least some of these issues, by concurrent simplification of the spatial structure and drastic increase in temporal resolution of the simulation. Our pilot study has shown that even a very simple, 1D (temporal) toy model of such phenomena is a fruitful approach  \citep{2018POBeo..98..121D}, and has the upper hand over the similar approach realized with 2D cellular automata  \citep{2010POBeo..90...85V,2010arXiv1007.2774B,galera_galanti_kinouchi_2019}, since the spatial aspect of the model is greatly simplified. This enables much better temporal resolution of the simulation that can model phenomena with $\sim10^4$ yr characteristic timescale during the lifespan of the Galactic disk. This is comparable to  the historical span of human civilization on Earth. On the other hand, the  temporal aspects of evolution of life can be studied against our empirical knowledge of the Earth's fossil record, while even the highly successful contemporary  exo-planet studies have yet to figure out the empirical signatures of life on other planets. Even  with multiple modern panspermia studies \citep[e.g.][]{2004MNRAS.348...52W,2016MNRAS.455.2792L}, it is still highly contentious how efficient naturally occurring panspermia is in the Galaxy \citep{Gordon_Hoover_2007,mcnichol2012we}. On the other hand, the efficiency of {\bf directed} panspermia is not in doubt, in contrast to the frequency of the technological civilizations capable of undertaking it, which remains unknown. Therefore, it makes sense to try to model both jointly. Also, the model presented is rather conservative in that it employs the conventional notion of habitable planets and fairly gradual evolutionary processes (even if punctuated by occasional catastrophes), and it does not include a type of more radical transformative processes, in particular postbiological evolution of extraterrestrial intelligence \citep[e.g.][]{dick2003cultural}.

This paper is organized as follows. The description of code, its structure, and phenomenology included is given in Section 2. Section 3 contains the description of simulations, with the computing resources employed. The main results are presented in Section 4, while the discussion and prospects for further work are given in the concluding section. 

\section{Method}

Unlike standard cellular automata approach, where spatial influence is implemented through the interaction of the cell with its neighbours, our model is simplified so that probability of colonizing (or possible panspermia seeding) of a particular spatial site is dependent on the total number of colonization capable entities (sites that have sufficiently evolved to be able to colonize other sites) in a given time step of the simulation. From the technical point of view, this can be regarded as a cellular automaton where each cell has a neighbourhood comprised of all other cells in the simulation. While this approach limits the inference on for example, modes of spatial spreading, such as simple spherical expansion vs. resource-driven percolation schemes \citep[e.g.][]{galera_galanti_kinouchi_2019}, the benefits from higher resolution of temporal scales are evident.  The main features of our model are as follows:

\begin{itemize}

\item The habitable sites in the Galaxy are represented as entities that have a particular state which evolves over time.    
\item Each site can evolve through four discrete states designated as: 0 -- no life, 1 -- simple life, 2 -- complex life, 3 -- technological civilization. 
\item Each site evolves according to probabilistic evolution rules with fiducial timescales representing the Earth's fossil record and planetary formation rate. During the course of the simulation, the sites are gradually activated to state 0 as per distribution of the Earth-like planets' formation rate from { \citet{2001Icar..151..307L}}.
\item Each cell in state 3 can colonize only one cell in state 1 or 2 per time-step, according to a Poisson-like probability distribution.   
\end{itemize}
We do not enter into complex and partially philosophical issues such as what exactly constitutes a technological civilization. Instead, we have adopted pragmatic approach of classic studies such as \citet{1981Icar...46..293N} that technological civilization is an entity capable - unless destroyed by an existential catastrophe - of colonizing nearby habitable planetary systems, sooner or later. We also used the terrestrial values in the present proof-of-concept model as the starting point, since the Copernican assumption has served us well in both science in general and astrobiology in particular (for an extended discussion of the latter see \citeauthor{cirkovic2012astrobiological}, \citeyear{cirkovic2012astrobiological}).

\subsection{Code description}

At the beginning of the simulation there are no active sites. As the simulation run progresses, the 1000 simulated  sites are activated to state `0' according to probability density distribution representative of cosmic Earth-like planet formation rate from \citet{2001Icar..151..307L} \citep[for other works see also][]{2016ApJ...833..214Z}. We used only the most recent 10 Gyr of this distribution and normalized it in such a way as to activate all sites (in a probabilistic manner) during our simulation time span of 10 Gyr. The frequency of site activation peaks at $\sim 1.5$ Gyr after the beginning of the run and than falls off with a shallower slope towards the end of the simulation. Also, once activated, each site is assigned the main sequence lifetime of $10^{10}m^{-2.5}$, where $m$ is the stellar mass selected randomly with uniform probability from $[0.6-1.3]\,$M$_\odot$ mass interval. If the time from the site activation exceeds the assigned main sequence time the site is removed from the simulation.   

The modelled transitions are presented in Figure 1. Each site might preserve its state to the next time step, sites in 0-2 states can evolve to +1 higher complexity just by internal evolution, state 1 and 2 sites might be colonized to state 3, while only state 3 sites can be degraded to state 2 by catastrophic events (natural or artificial). The transitions between the states are implemented in a similar manner as in \citet{2012OLEB...42..347V}. The transition probabilities in that work (arranged in a matrix for the sake of formalism and easier reference) have been calculated from the timescales describing those transitions. The main difference is that this work uses Epanechnikov instead of a Gauss kernel to calculate transition probabilities from the parameters of the relevant transition timescales (Table 1). The active site in state $i$ can evolve to state $j$ according to Epanechnikov kernel \citep{Epanechnikov}, $E(u) = 0.75(1-u^2)$, where $|u|\leq1$. Epanechnikov kernel is much less computation intensive than widely used Normal distribution kernel. {  As explicated in our recent pilot study concerning the topic of this work \citep{2018POBeo..98..121D}, at the beginning of each run, each of the simulated sites is assigned a set (one for each possible transition) of random variates from a uniform probability distribution in $(0-1)$ interval ($f$)}. The state of the site is changed  if the resulting cumulative density function at a time $t$ has higher value than this pseudo-random number \citep[we used a pseudo-random number generator SFMT, see][]{saito2008simd}: 

\begin{equation}
E_{cdf}\left(\frac{ t -\tau}{\sigma}\right) > f,
\end{equation}
where $t$ is time that a site have spent in state $i$ since activation or the last transition, if applicable, $\tau$ is a characteristic time (mean value of kernel) for transition $i\rightarrow j$ where $j$ is  the next state  and $\sigma$ is standard deviation, that is, width of the kernel. When the transition occurs for the particular object, a random variate for that particular transition is reinitialized. The above model is used for evolution-type transitions, where site evolution is not to be influenced by spatial spreading agents, such as colonization or panspermia. A similar equation is adopted for the calculation of probability for colonization-induced transitions:
\begin{equation}
E_{cdf}\left(\frac{ t_\mathrm{3} -\tau_\mathrm{col}}{\sigma_\mathrm{3}} \right)   > f,
\end{equation}
 where  $t_\mathrm{3}$ is  time that object have spent in state 3 (up to the current time step), $\tau_\mathrm{col}$ is the characteristic time of colonization while it's standard deviation is $\sigma_\mathrm{3}$.  When the state 3 object reaches a time step where  the above condition is fulfilled it makes a colonization attempt and it's $t_\mathrm{3}$ gets reset to zero regardless of colonization success. The colonization attempt is considered a failure if there are no objects in state 1 or 2 (at a particular time step), which is highly improbable. If there are objects in state 1 or 2 then the following conditions are tested:
 \begin{eqnarray}
 \frac{N_1}{N_\mathrm{a}} < f_1,\\
 \frac{N_2}{N_\mathrm{a}} < f_2.
 \end{eqnarray}
The total number of objects in state 1 and 2 are represented with $N_1$ and $N_2$, respectively, while  $N_\mathrm{a}$ is the total number of  activated objects from the beginning of the simulation run. The above conditions are tested for each colonizing (state 3) object at a given time step. Two separate random variates $f_1$ and $f_2$ are generated each time the conditions are tested. If the first condition is true, then the  randomly selected object in state 1 is colonized. Otherwise, if the first condition is false, then the second condition is tested. If the second condition is true, the randomly selected state 2 object is colonized. In the case when the second condition is false (i.e. both conditions are false), then the corresponding colonization attempt is considered a failure. When the object in state 1 or 2 gets colonized, the random variate of the corresponding evolutionary $1\rightarrow 2$ or $2\rightarrow 3$ transition for that object is also reinitialized. While this might not be of practical importance for objects in state 1, objects in state 3 can be catastrophically degraded back to state 2.

The catastrophic aspect of the model is implemented with a separate  $3\rightarrow 2$ transition as a Poisson-like process. In each time step the object in state 3 must satisfy:
\begin{equation}
\frac{t_\mathrm{res}}{\tau_\mathrm{cat}} < f_\mathrm{cat},
\end{equation} 
where $t_\mathrm{res}$ is a simulation time step, $\tau_\mathrm{cat}$ is a characteristic time scale for catastrophic events and  $f_\mathrm{cat}$ a uniform random variate generated each time the condition is tested. Otherwise, the object is degraded to state 2. Table 1 summarizes transitions and their parameters.
\begin{table}
\caption{Parameters of the transitions in our model. From left to right: transition designation, type, characteristic time, standard deviation. }
\centering
\begin{tabular}{c c c c}\hline\hline
$i\rightarrow j$ & Type&$\tau$ & $\sigma$ \\
&&[yr]&[yr]\\\hline
$0\rightarrow 1$ &  evolution&$5\times10^8$ & $3\times10^8$ \\
$1\rightarrow 2$ &  evolution&$3\times10^9$ & $1\times10^9$ \\
$2\rightarrow 3$ &  evolution&$6\times10^8$ & $1\times10^8$ \\
$3\rightarrow 2$ &  catastrophic&$[10^4, 10^{10}]$ & $0.1*\tau$ \\
$1\rightarrow 3$ &  colonization&$[10^4, 10^{10}]$ & $0.1*\tau$ \\
$2\rightarrow 3$ &  colonization&$[10^4, 10^{10}]$ & $0.1*\tau$ \\
\hline
\end{tabular}
\end{table}

For the characteristic times from Table 1,  for the emergence of life and its evolution to complex forms are according to \citet{dodd2017evidence} and \citet{maloof2010earliest}, respectively. Finally, we used a figure of 600 million years for the development of intelligent life and civilization, roughly following the Copernican assumption about typicality of evolution of complex life and intelligence on Earth. The catastrophic timescale that represent high risk events such as close gamma-ray burst, supernovae, and asteroid impacts is variable in the interval indicated in Table 1. Independent of catastrophic transition, the two timescales for the colonization transitions are varied together (have equal values in each simulation). We have neglected colonization rates to sites in state 0, in order to avoid seeding of planets incapable of supporting life at given time. This makes the model more conservative and more in line with the conventional scenarios of colonizing planets possessing at least some biotic resources.

\begin{figure*}
\centering
   \includegraphics[width=8.5cm]{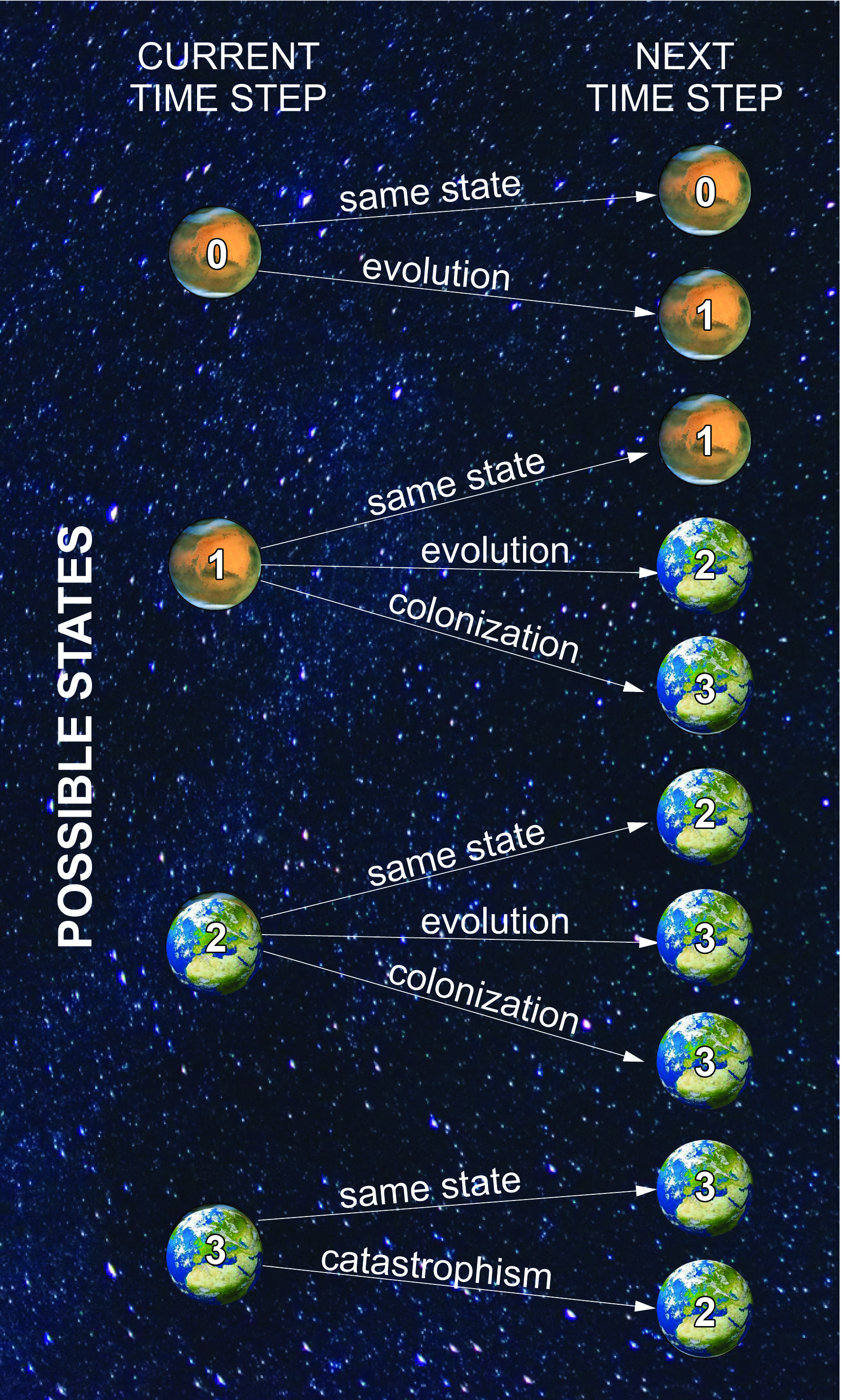}\includegraphics[width=8.5cm]{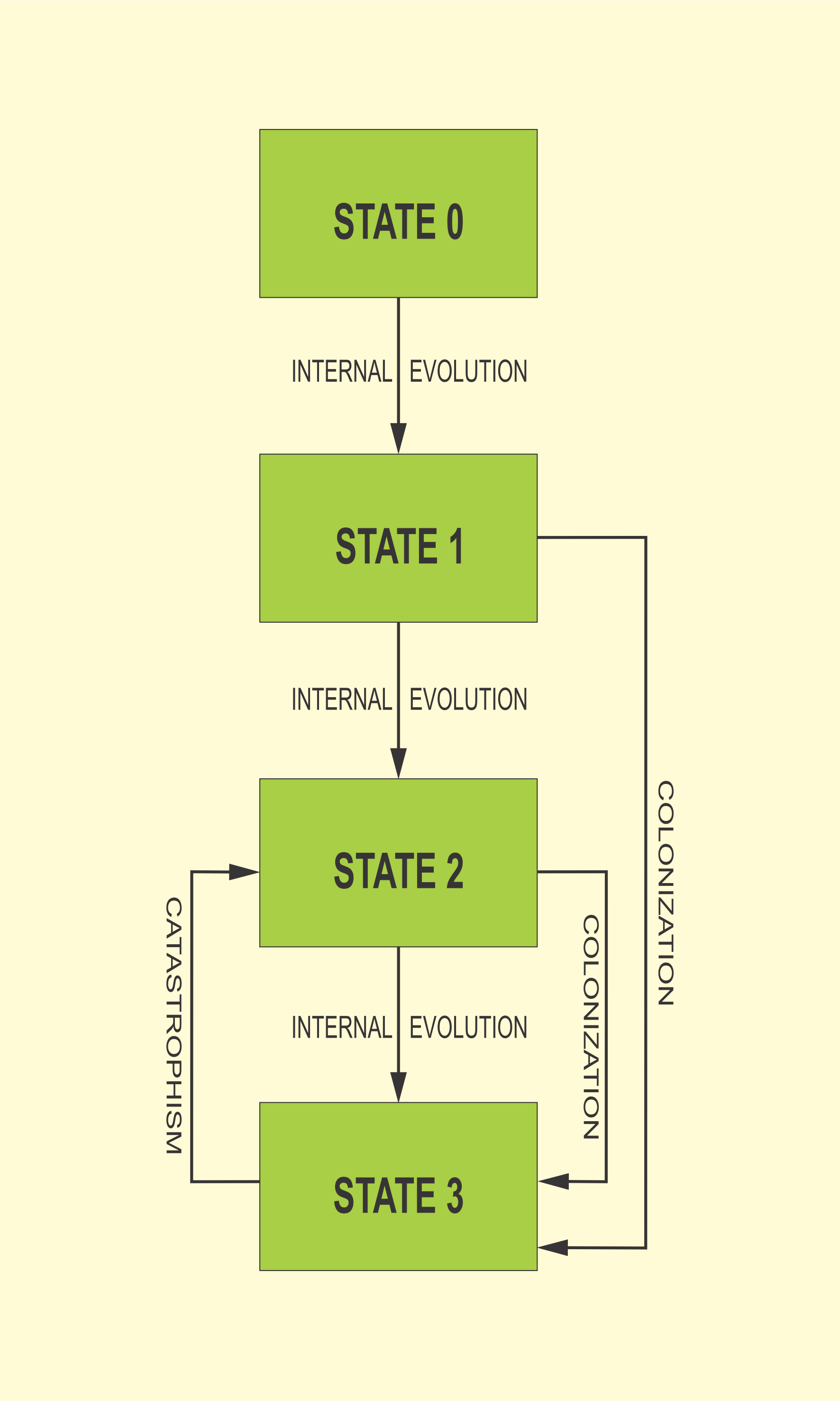}
\caption{Modelled  transitions with a symbolic scheme of states (we assume that a Mars-like planet could have been habitable for simple life in at least some epochs of its history).}
\end{figure*}

\subsection{Simulation technical details}
The simulations were run on PARADOX III super-computer at The Institute of Physics Belgrade. PARADOX III consists of 106 computing nodes. Each node has 16 CPU cores (2x8) Sandy Bridge Xeon 2.6GHz, with 32 GB RAM memory. All nodes are interconnected by the QDR InfiniBand network. A great deal of the required computation was also carried out at the Bibliotheca Alexandrina High-Performance Computing machine, BA-HPC C2. BA-HPC C2 consists of 98 servers (each with 128 GB of RAM memory), with a total of  1968 CPU cores and 288 TB of storage. 

Each simulation covers the time span of 10~Gyr in steps of 10~kyr.  The input timescales for colonization and catastrophism were different for each simulation. They were selected from an interval of $10^4$ to $10^{10}$~yr on exponential scales in increments of 0.05. This gives a total of 14641 combinations of input parameters which is also the total number of simulations performed  (and the total number of the output files). We ran one simulation per core. A pseudo random number generator is seeded with the number of the simulation, that is, from $(0, 14641)$ interval.  The results of each simulation are an average of 100 runs. The described models for colonization and catastrophism accounted for the most of the computational resources. The runs with a significant number of state 3 objects took up to a $10^4$ s of wall-clock time, while some of the fastest were completed in a matter of $7 \times 10^2$ s. The downside of the sequential output is that it is not feasible to perform `atemporal' analysis of the model (e.g. to calculate the number of sites at a fixed state over all snapshots). 

\section{Results}

We present the results of our simulations in Figures 2-4. Figure 2 shows the colour coded plot of the average total number of sites in state 3 at the end of the simulation (at 10 Gyr) for each set of varied characteristic timescales. For the greater part, the plot surface shows the number of objects to be below the middle point of the colour scale (the indicated mean value is below 450). This apparent dominance of catastrophism is also evident in the prominent diagonal feature lying below the $\tau_\mathrm{cat} = \tau_\mathrm{col}$ line. The middle value of the colour bar settles at $\approx10^8=\tau_\mathrm{cat}=\tau_\mathrm{col}$. The light blue to orange region in the upper right corner of the plot with $\tau_\mathrm{cat}=\tau_\mathrm{col}  > 10^8$ is not dominated by catastrophic events. There, even if $\tau_\mathrm{cat}<\tau_\mathrm{col}$, it is possible for the number of objects in state 3 to be above the average. The gradient of colour coded values in this part is also smaller than in the catastrophism dominated region. The area with $\tau_\mathrm{col} < 10^5$ and $\tau_\mathrm{cat} < 10^6$ shows some fluctuations.

In Figure 3 we plot how the average number of sites in each state changes with time for a selected set of variable input parameters. The standard deviation of curves from Figure 3 is given in Figure 4. While Figure 2 is useful for discussion of the emerging `habitability regimes', Figures 3 and 4 are more suitable for grasping the underlying mechanics of the evolution timescales and their relation to different evolution scenarios.

Each data point in Figures 2 and 3 is the mean value of 100 runs of the simulation. The first row of panels in Figures 2 and 3 shows that catastrophism overwhelms the colonization and there are almost none of the objects in state 3 throughout the whole simulated time span. The standard deviation for all emergent curves is at a level of approximately ten which implies no sudden changes and a steady evolution. 

This is very similar to the third row of panels with colonization as a dominant agent. The main difference however is a region of instability between $3$ and $4$ Gyr (left most and middle panel). Once the colonization starts it takes about $0.5-1$ Gyr (for $\tau_\mathrm{col} = 10^5$ up to $\tau_\mathrm{col} = 10^7$, respectively) to colonize most of the objects in states 1 and 2. Once the equilibrium with catastrophism is evident in the middle panel of this row and the number of objects in state 2 is approximately ten until the end of the simulation. The panel on the right hand side shows no such behaviour and is very similar to the panels in the first row, except for existence of a significant number of state 3 objects, mostly at the expense of the population in state 2. Similar curve shapes, but with lesser number of objects in state 3 are emergent in the middle row in central and the right hand side panel, while the left most panel in this row shows a typical signature of dominant colonization.
  
A somewhat larger fluctuations, both in the numbers of state 3 objects and in their standard deviation is evident in the middle row (when compared to the third row). The larger values for catastrophic timescales give a smooth appearance of the curve while smaller values suppress the appearance of state 3 sites. The small number of state 3 sites in the right-hand side and in the middle panel of the middle row in Figure 3, approximately ten, shows a slightly larger relative spread and a non-stable behaviour (Figure 4). Apart from being of little value for making the quantitative conclusions from the statistical point of view it also lacks credit from a methodical view point. \citet{cirkovic2018great}, argues that even the discovery of a single to a few advanced civilizations would have a little power in resolving Fermi's paradox in its stronger forms.  The similar standard deviation curve for states 1 and 3 in colonization dominated cases (at $3-4$ Gyr time interval) is due to the fact that onset of colonization happens at a rather stochastic and  unpredictable pace, because small variations in the number of state 3 objects at the beginning of this phase can proliferate fast to much larger differences. This directly imprints on the state 1 curve since it is subjected to colonization at a pace dictated by the state 3 curve. A similar process should hold for state 2 curves, but this is not obvious from the presented plots as state 2 does not have enough time to build up in numbers as state 1 does.   
 
\begin{figure*}
\centering
\includegraphics[width = 17cm]{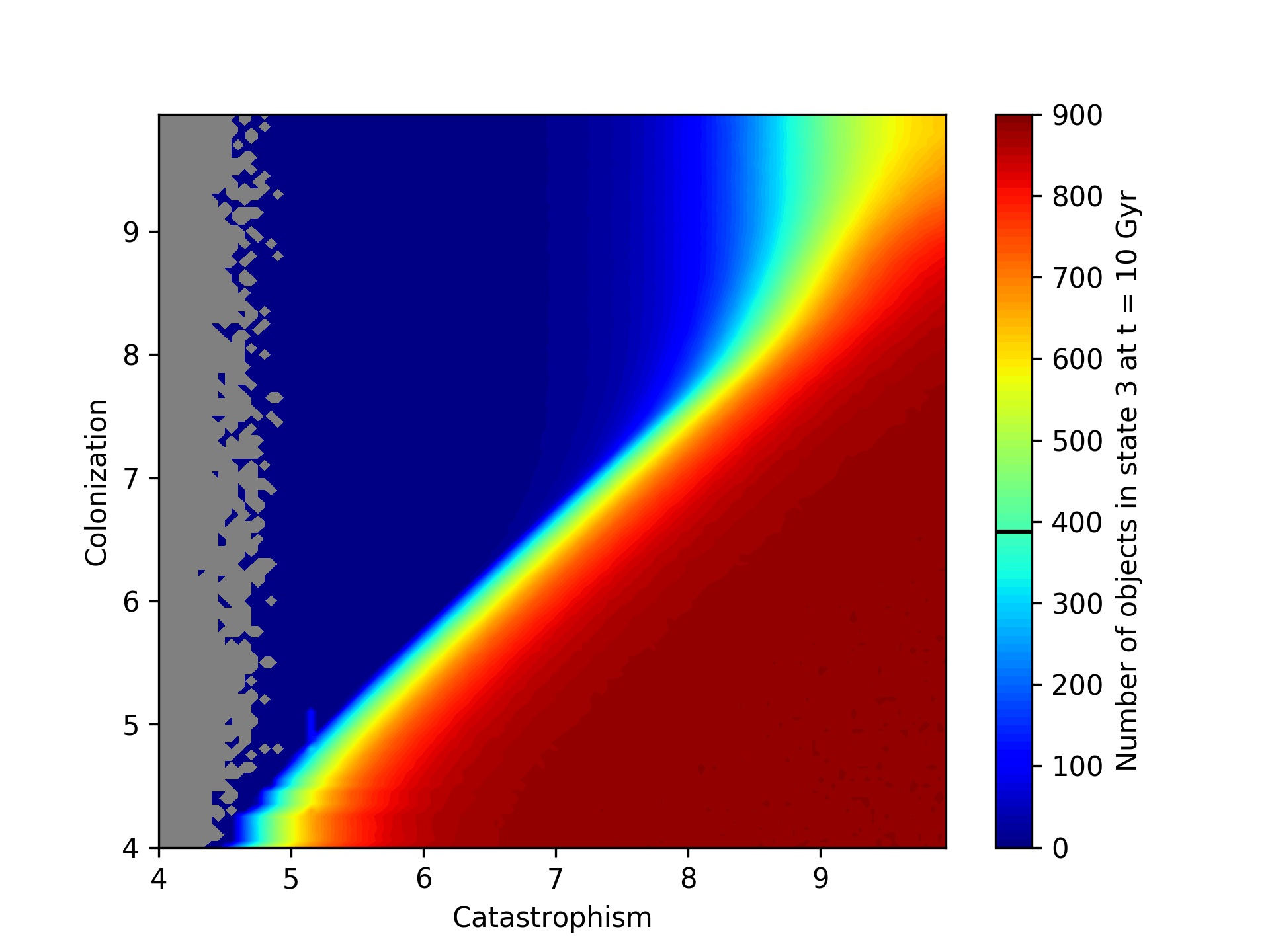}
\caption{Dependence of the number of sites in state 3 as a function of timescale for catastrophism and colonization at the end of the simulated time span (10 Gyr). Axes are designated with decadic logarithm of time in years. The grey colour represents pixels with exactly zero number of sites in state 3, while the numbers values higher  than zero are colour-coded as displayed. The mean value of the plotted matrix is $391.90$ and is indicated with the black dash on the colour bar.}
\end{figure*}

\begin{figure*}
\centering
\includegraphics[width=17cm]{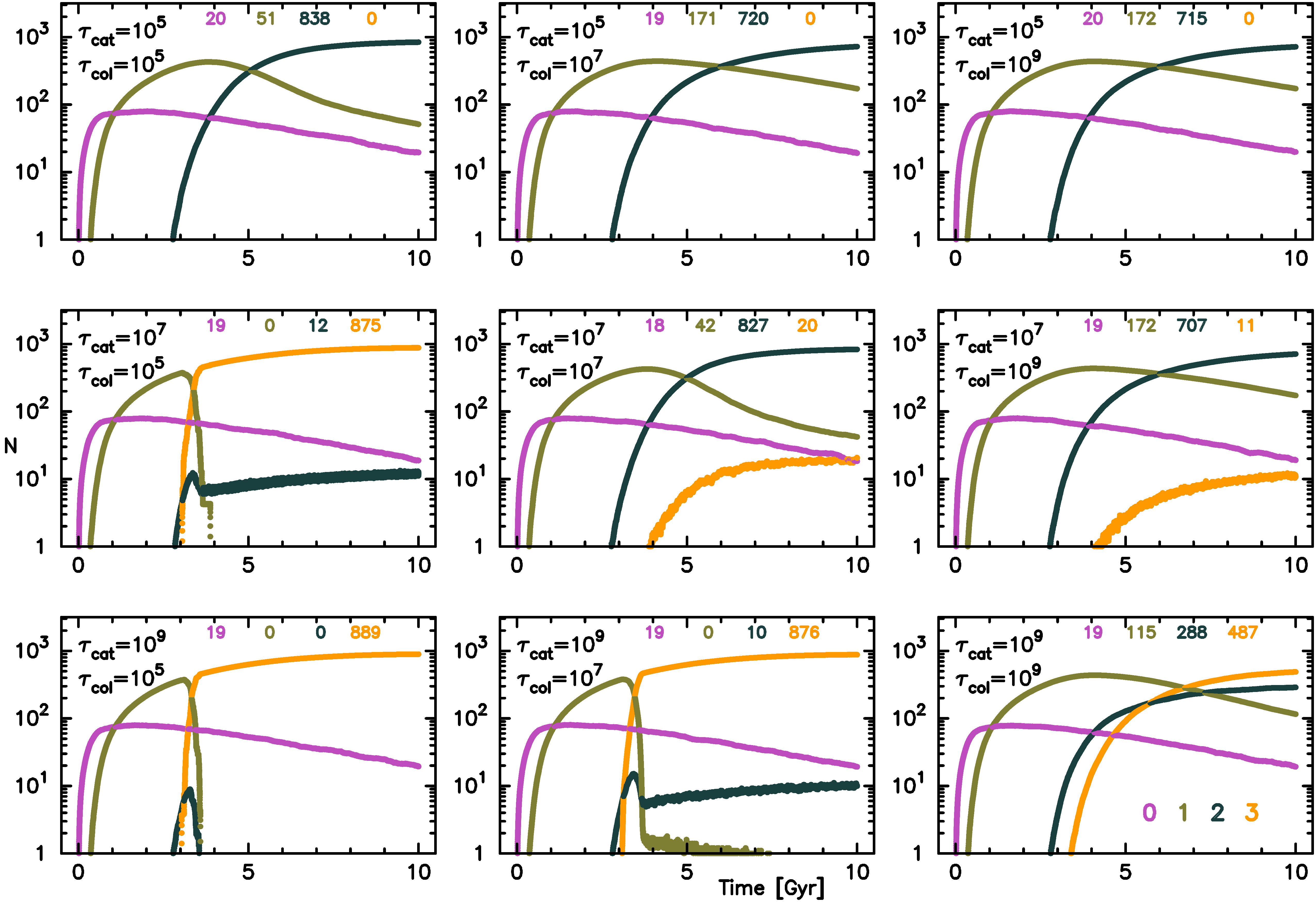}
\caption{Number of cells in each state (colour-coded as listed in bottom right panel) as a function of time. Different plots means different initial conditions for colonization timescale ($\tau_\mathrm{col}$) and catastrophism ($\tau_\mathrm{cat}$). The number of sites in each state after 10 Gyr is indicated at the top of each panel.}
\end{figure*}

\begin{figure*}
\centering
\includegraphics[width=17cm]{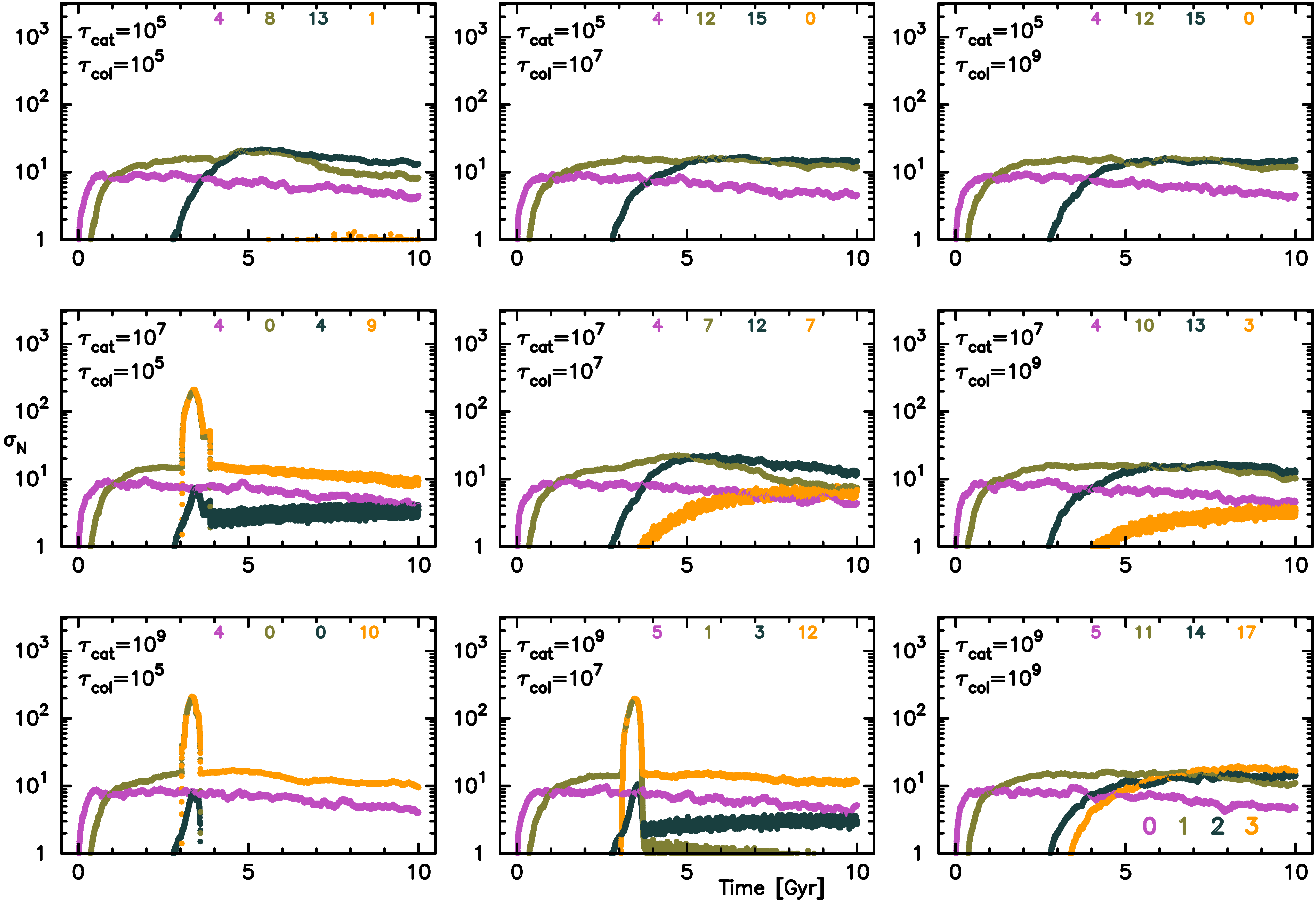}
\caption{Similar plot as in Figure 3, except that it shows the standard deviation of the number of cells in each state.}
\end{figure*}

\section{Discussion}    

The larger part of the plot surface in Fig. 2 is grey or dark-blue-coloured, indicating that catastrophic events might play a dominant role in shaping the parameter space for the evolution of life. Even only in a mild catastrophic regime, with no $3\rightarrow 1$ transitions. This might be a consequence of the $1\rightarrow 2$ evolution characteristic time of 3 billion years. If life took very "little" time to appear on Earth it is somewhat contradictory that it required another 3 billion years to make complex forms. Either some form of astrobiological regulation (possibly catastrophic) might be at play or it can simply be attributed to the enormous fitness of extremophile single cell life. This fits well into our current empirical inventory: there are many planets out there, but still no perception of the advanced exo-civilizations. The dominance of catastrophism is also obvious from plot features extending along the colonization scale, which show an independence of those features from colonization and strong dependence on selected catastrophic times. Another interesting interpretation would be that cosmic catastrophes followed by mass (but not total) extinctions may indeed have a positive effect on evolution as a whole (through evo-devo mechanisms, adaptive radiations similar to the Cambrian explosion, etc.), as has been speculated, for instance, by the `rare Earth' theorists \citep{2000rewc.book.....W}, following an early idea of the physicist John Cramer \citeyear{1986Cramer}. With that interpretation, we would rather expect a U-shaped dependence, even if skewed, of the frequency of complex life on the rate of occurrence of catastrophic events.

Our model of transitions is very conservative as far as colonization is concerned, at least when compared with catastrophism. From Figure 1, state 3 can be pumped with evolution from state 2 and colonization from states 1 and 2. At the same time it is drained only by catastrophism to state 2. The overall compendium of empirical evidence from astrophysics, biology, geology, sociology, philosophy,  and even anthropology) implies that life-hazardous events have occurred in the past and are a probable threat in the future. Even our present existence on Earth, in the light of the probabilistic anthropic selection effect \citep{cirkovic2010anthropic}, is consistent with a significant underestimate of hazardous event frequency. On the other hand, we have no such empirical background in estimating relevant colonization timescales; at least without significant extrapolation to future technologies. Enforcing a certain level of handicap to interstellar colonization over naturally- or technologically-induced catastrophes should give more restrictive but reliable results as far as outlining the parameter space regions with a large number of state 3 sites is concerned. For this purpose, state 3 objects were restricted to make a colonization attempt towards only one possible target during the $\tau_\mathrm{col}$. In addition, the success of colonization depends also on the relative number of the available targets. It is evident from Figure 2 that at the onset of colonization the rate  of making the state 3 objects has a rapid increase and then gradually decreasing as the number of state 3 objects becomes larger. 

Even with such restrictions the $\tau_\mathrm{cat}=\tau_\mathrm{col}  > 10^8$ region is dominated by a high number of state 3 objects. It appears that this is the consequence of setting the $2\rightarrow3$ characteristic evolution time to $3\times10^8$ yr. This implies that the shape of distinction pattern between colonization- and catastrophism- dominated regions is primarily determined by the ability of the habitable sites to recover from the catastrophic events. With the recovery pace being larger than the frequency of hazards, the state 3 objects clearly dominate the scene. Also, the steeper slope of this pattern in the $\tau_\mathrm{cat}=\tau_\mathrm{col}  > 10^8$ region indicates that there is less dependence on the colonization time scale, even for points in excess of 500 state 3 objects at the end of the simulation. Here, the significant number of state 3 objects should not be contributed to colonization but rather to evolution and increased life time because of the larger $\tau_\mathrm{cat}$.

The panels on the main diagonal in Figure 3, $\tau_\mathrm{cat}=\tau_\mathrm{col}$ show a very interesting behaviour of state 1 curve. With respect to the results presented in Figure 2, the catastrophism-dominated part (with both timescales at $10^5$ or $10^7$ yr) shows a convex shape of state 1 curve at epochs later than $\sim5$ Gyr, resulting in a comparable number of state 1 and state 0 objects at the end of the simulation. Figure 4 shows that, although in Figure 3 there are no state 3 objects for $\tau_\mathrm{cat}=\tau_\mathrm{col}=10^5$ yr, objects in state 3 do appear during the simulation run. The shape of the state 1 curve might be explained by the colonization action of state 3 upon state 1. Some of the state 3 objects manage to colonize state 1 objects before they are catastrophically degraded to state 2. This effect reduces towards later epochs since the number of available state 1 targets gets smaller. Since the state 2 curve does not show similar features, likely because of also being catastrophically populated back from state 3, the shape of the state 1 curve might be indicative of the colonization action even in the absence of detecting (or existence) of colonizing objects themselves at the present epoch. The quantification of this effect requires further examination and fine-tuning  of the colonization model.

With regard to our current ignorance of exo-civilizations, the red-coloured part of the parameter space in Figure 2 is not of particular interest for better understanding of the Fermi's paradox, while the yellow and light-blue regions might be very significant. In other words, all the regions of the parameter space with a very small number of state 3 objects, especially the sub-parts with smaller colonization timescales (i.e. $\tau_\mathrm{col}<10^8$ yr). The colour bar in Fig. 2 indicates that a number of colonized sites can reach up to a $20\%$ of the total number of sites with a very steep gradient. This implies that the conventional (and perhaps naive) view of gradual, melioristic evolution towards increasing complexity -- which is incompatible with our observations -- should be reconsidered. Clearly, the higher resolution sampling of the parameter space is of particular importance for measuring this gradient and relating it to the adopted evolutionary timescales.

In addition to an earlier study by \citet{2011AsBio..11.1041S}, a recent surge of works \citep{2018LSSR...19...13J,2018A&C....23..141S, 2018OJAp....1E...4K,2017MNRAS.471.4628R,2017Ap&SS.362..146K,2015ApJ...814...91B} demonstrate the classification possibility of  the discovered population of exoplanets according to their habitability. The primary motivation for these works is the advance in the field of exo-life search. However, these works also offer a prospect for advancing our understanding of Fermi's paradox. A detailed comparison of the relative abundance of objects in different states (such as the curves in Figure 3) and habitability classification of exo-planets might introduce an abundant source of empirical data to further constrain the parameter space of possible evolutionary timescales.

\section{Summary}

We have presented the results of our study that used a simplified version of the spatial spreading of life in the Galaxy in order to achieve higher computational efficiency for the purpose of mapping the colonization vs. catastrophic timescales parameter space. The other relevant timescales for the life evolution (biological) are tentatively taken to be adequately represented by the Earth's fossil record, as per the Copernican principle. The emerging shape of the number of highly evolved sites shows lower than average values for the greater part of the examined parameter space, indicating the dominance of catastrophism over colonization. This fits well into our current perception of exolife and as our results have demonstrated,   indicates  a possibility of a fine-tuning solution to Fermi's paradox. However, further understanding and improvements of the present model are required. Among possibilities to be considered in the course of the future work, there are both `conservative' (better theoretical models of the habitable planets' ages, and different averaging procedures) and `radical' ones (more different discrete sites, and a wider Poissonian distribution of the transition timescales, taking into account correlated artificial disasters like large-scale wars or pandemics). In any case, the present proof-of-concept study clearly shows that discrete models of this kind are capable of capturing some of the complexity of astrobiological evolution.

\begin{acknowledgements} 
We thank an anonymous referee for an extensive and insightful review that has greatly improved the overall content of the paper. Fruitful discussions with Srđa Janković, Milan Stojanović, Anders Sandberg, and John Smart have enormously contributed to the ideas presented in this paper. The authors acknowledge financial support from the Ministry of Education, Science and Technological Development of the Republic of Serbia -- VĐ through the project \#ON176011 ``Dynamics and kinematics of celestial bodies and systems'',  BV and MM{\'C}  through the project \#ON176021 ``Visible and invisible matter in nearby galaxies: theory and observations''. Numerical simulations were run on the PARADOX supercomputing facility at the Scientific Computing Laboratory of the Institute of Physics Belgrade as a part of the MALAHIT project and on the Bibliotheca Alexandrina High-Performance Computing facility within the VI-SEEM initiative project CAT-ICE.  In this regard, the authors wish to express their special gratitude to Antun Balaž, Aleksandar Bogojević, Aleksandar Belić, Dušan Vudragović, and Petar Jovanović.  
\end{acknowledgements}

\bibliographystyle{aa} 
\bibliography{djosovic_ref.bib}

\end{document}